\documentclass{article}

\usepackage{PRIMEarxiv}
\usepackage{comment}
\usepackage[utf8]{inputenc} % allow utf-8 input
\usepackage[T1]{fontenc}    % use 8-bit T1 fonts
\usepackage{hyperref}       % hyperlinks
\usepackage{url}            % simple URL typesetting
\usepackage{booktabs}       % professional-quality tables
\usepackage{amsfonts}       % blackboard math symbols
\usepackage{nicefrac}       % compact symbols for 1/2, etc.
\usepackage{microtype}      % microtypography
\usepackage{lipsum}
\usepackage{fancyhdr}       % header
\usepackage{graphicx}
\usepackage{algorithmic}
\usepackage{algorithm}% graphics
\graphicspath{{media/}}     % organize your images and other figures under media/ folder

\title{Enhancing Electrocardiogram Signal Analysis Using NLP-Inspired Techniques: A Novel Approach with Embedding and Self-Attention
}

\author{
  Prapti Ganguly, Wazib Ansar, Amlan Chakrabarti \\
  A.K.Choudhury School of Information Technology \\
  University of Calcutta \\
  Kolkata, India\\
}

\begin{document}
\maketitle

\begin{abstract}
A language is made up of an infinite/finite number of sentences, which in turn is composed of a number of words. The Electrocardiogram (ECG) is the most popular noninvasive medical tool for studying heart function and diagnosing various irregular cardiac rhythms. Intuitive inspection of the ECG reveals a marked similarity between ECG signals and the spoken language. As a result, the ECG signal may be thought of as a series of heartbeats (similar to sentences in a spoken language), with each heartbeat consisting of a collection of waves (similar to words in a sentence) with varying morphologies. Just as natural language processing (NLP) is used to help computers comprehend and interpret human natural language, it is conceivable to create NLP-inspired algorithms to help computers comprehend the electrocardiogram data more efficiently. In this study, we propose a novel ECG analysis technique, based on embedding and self attention, to capture the spatial as well as the temporal dependencies of the ECG data. To generate the embedding, an encoder-decoder network was proposed to capture the temporal dependencies of the ECG signal and perform data compression. The compressed and encoded data was fed to the embedding layer as its weights. Finally, the proposed CNN-LSTM-Self Attention classifier works on the embedding layer and classifies the signal as normal or anomalous. The approach was tested using the PTB-xl dataset, which is severely imbalanced. Our emphasis was to appropriately recognise the disease classes present in minority numbers, in order to limit the detection of False Negative cases. An accuracy of 91\% was achieved with a good F1-score for all the disease classes. Additionally, the the size of the model was reduced by 34\% due to compression, making it suitable for deployment in real time applications. 
\end{abstract}

% keywords can be removed
\keywords{PTB-XL \and ECG \and attention \and embedding \and compression}

\section{Introduction}
{C}{ardiovascular} diseases continue to be the leading cause of death across the globe and automated diagnosis of the same has become an imperative need of the moment for healthcare providers \cite{owidcausesofdeath}. The electrocardiogram (ECG) is a valuable and the most widely used diagnostic tool used for detecting such cardiovascular irregularities. It is a non-invasive procedure that records the electrical activity of the heart and  reveals the abnormalities in cardiac rhythms\cite{maron2014assessment},\cite{jin2018screening}. Recent developments in medical technology allow for the simultaneous collection of enormous amounts of ECG data, which needs processing and interpretation by medical experts. However, the process is time-consuming and expensive, as it necessitates the involvement of doctors with specialized expertise. This poses a huge challenge, particularly in villages and remote places, where the availability of medical facilities are limited with a very few medical practitioners.
 \par To provide a greater insight into remote monitoring of ECG data, a reliable automated ECG-analysis tool is required. Although numerous efficient single-lead ECG arrhythmia classification approaches have been put forth in \cite{jambukia2015classification}, \cite{kiranyaz2015real} and \cite{kachuee2018ecg}, single-lead ECG alone is insufficient to effectively diagnose a variety of heart conditions as discussed in \cite{baloglu2019classification}.  Each lead in a 12-lead ECG provides insights into the heart's condition by capturing electrical signal variations from different perspectives. Therefore, 12-lead ECG analysis has garnered increased attention from researchers as a typical clinical ECG assessment tool as described in \cite{liu2018multiple} and \cite{strodthoff2020deep}.

 \par The initial automated diagnostic approaches for ECG signals involve a combination of feature extraction and classification techniques. The most important aspect of these approaches is the extraction of discriminant information from original ECG data, also known as feature extraction. These extracted ECG features are then fed into a classifier to accomplish the task of signal classification. ECG signals are contaminated with noise due to baseline wander, motion artifacts of muscles, electro-myographic noise, instrumentation noise, etc, which makes it difficult to identify disease specific morphological features in the waveform \cite{MONDEJARGUERRA201941}. Several filtering approaches have been reported over the years to filter out the various noise present in the data. In \cite{9624454}, the authors demonstrated a wavelet-based de-noising technique for non-stationary signals utilizing an enhanced threshold function. In \cite{10.1007/978-981-16-7011-4_7} the authors demonstrated the use of a median filter to eliminate all conceivable artifacts, including baseline wander. Several other feature extraction techniques include empirical mode decomposition (EMD) \cite{7566382} and discrete cosine transform (DCT) \cite{iet:/content/journals/10.1049/el.2010.3191}.
 \par Traditional machine learning methods have been applied for the purpose of identifying cardiovascular diseases. Support Vector Machines (SVM) have performed strongly in ECG classification as shown in \cite{USHAKUMARI20211393} and \cite{MONDEJARGUERRA201941}, while demanding a huge computation cost. In \cite{9588215}, the authors have  compared the ECG anomaly detection accuracies of several classical machine learning classifiers, including SVM, Naive Bayes (NB), Random Forest (RF), K-nearest Neighbour (KNN), and Decision Trees (DT). XGBoost and Adaboost have been employed for classification in \cite{rezaei2021novel} and \cite{rajesh2018classification} respectively. The average sensitivity for the disease classes was reported to be less than 90\%, which suggests that many False Negative instances have been detected. The efficiency of these classifiers is highly dependent on developing discriminative feature extraction techniques and the relevance of the features used for classification.
\par Deep learning has demonstrated its value in numerous real-world applications including computer vision, speech recognition, and natural language processing due to its ability to capture complex patterns and dependencies within the data.  Researchers across various domains, including technology, healthcare, finance and automobiles, have embraced deep learning to ehnance their products, services and operations. These success stories have generated significant interest and motivation for further exploration of deep learning techniques. Deep neural networks (DNNs) have the capability to  automatically extract valuable features from the raw data thereby reducing the likelihood of of the classification accuracy being affected due to improper manual feature extraction. Additionally, DNNs are well-suited for handling large volumes of data without requiring extensive preprocessing. These advantages have contributed to the extensive use of DNNs in the analysis of ECG data. In \cite{al2018convolutional}, the authors suggested using continuous wavelet transform to turn one-dimensional ECG signals into two-dimensional images, followed by using convolutional neural network (CNN) to extract useful features from the images, producing a high accuracy for identifying abnormal ventricular beats. A 34-layer CNN network was constructed by Hannun et al. \cite{hannun2019cardiologist}, where the authors have classified twelve arrhythmias with a cardiologist-level F1 score of 0.837. Long-Short Term Memory (LSTM) network \cite{hochreiter1997long}, Recurrent Neural Network (RNN) \cite{gahane2022design}, and bidirectional LSTM network \cite{nam2022selective} are other popular deep learning models that have been adopted for ECG categorization. Because there may be lags of unknown duration between key events in ECG data, LSTM networks are well-suited for ECG time series classification, processing, and prediction as put forth in \cite{xu2020interpretation}. In \cite{tung2020multi}, the authors have used 2 convolutional layers along with 4 residual blocks and 2 BiLSTM modules to achieve a high accuracy of about 95\% in the classification of cardiovascular diseases.

\par NLP models to classify ECG signals was first employed in \cite{MOUSAVI2021105959}, where the authors followed a conventional NLP technique of creating a vocabulary, wave embeddings and finally training the model using both CNN and RNN techniques. 
Conventional NLP approaches
initially use clustering to find groups of similar
signals. To classify signals, a vocabulary is generated from
the available signals. The test signal is then compared to the
vocabulary to determine its similarity. Hence, these approaches
limit the size of the vocabulary beforehand. However, our
approach is novel since each signal is treated as a new entity in the vocabulary.

\par Our contributions in this paper are enumerated as follows:

\begin{enumerate}
    \item A novel approach is proposed where each new waveform in the signal is treated as a distinct entity in the vocabulary. This creates the provision for the inclusion of unseen signals, which might have not been encountered before, in the vocabulary.
    \item Compression of input data is carried out and passed on to the embedding layer for classification. This leads to reduced memory consumption as there is a significant reduction in the model size while maintaining a consistent accuracy.
\end{enumerate}

The rest of this article is organized as follows. Section \ref{Rel_w} gives a brief literature review of the previous works in this domain. Section \ref{dataset} gives an insight into the dataset used for this research. Section \ref{method} describes the methodology used in this study. The experimental design and hyperparameters are discussed in Section \ref{hyper}. Evaluation of the results is investigated in Sections \ref{results} along with the ablation studies elaborated in Section \ref{hyper}. The article concludes with Section \ref{conclusion}.

\section{Related Works}
\label{Rel_w}
Prior to presenting the suggested TAE-CLSA framework, background survey on attention mechanisms, NLP techniques and CNN used for ECG classification has been provided.
\subsection{Attention mechanism}
ECG signals often vary in length due to differences in heart
rate or length of recordings. Moreover, some sections of the
signal may contain more indicative features than the rest of
the signal. Attention mechanisms allow the model to focus
on informative segments while ignoring irrelevant or noisy
parts [26]. It helps in localizing the abnormalities by assigning
higher attention weights to the relevant regions. This allows
the model to emphasize the important parts of the signal and
improve the accuracy of anomaly detection. A Deep neural network (DNN) structure that integrates a residual convolutional network with an attention mechanism designed for the purpose of identifying anomalies in ECG signals has been studied in \cite{liu2019automatic}. Some of the disease classes achieved an F1-score about 90\% while the rest of the disease classes achieved a score of just about 70\%. In \cite{fu2020hybrid}, the authors have incorporated multi-lead attention (MLA) mechanism into a framework consisting of a convolutional neural network (CNN) and bidirectional gated recurrent unit (BiGRU), denoted as MLA-CNN-BiGRU, to identify myocardial infarction (MI) using 12-lead ECG data. Weights are assigned to each lead by the multi-lead attention mechanism, according to their contribution in detecting MI. The CNN module takes advantage of the relationships between leads' attributes to extract discriminatory spatial features, while, each lead's key temporal properties are extracted via the BiGRU module. The model succesfully detects myocardial infarction with decent accuracy and an average F1-score of 87\% . A spatio-temporal attention-based convolutional recurrent neural network (STA-CRNN) capable of extracting representative features from both spatial and temporal axes has been designed in \cite{zhang2020ecg} to classify eight types of arrhythmia and normal rhythm with an average F1 score of 0.835. In \cite{WANG2021106006}, a 33-layer CNN architecture with a non-local convolutional attention block module(NCBAM) has been developed. The CNN module extracts the local spatial and channel properties, and the non-local attention block captures the long-distance dependencies. Finally, a learnt matrix is utilized to combine the two blocks, yielding an F1 score of 0.85 on the PTB-XL database. The architectures of attention modules in different places are always the same in a network with artificially built attention modules, although the same architecture may not be suited for all positions. As a result, it is more logical to use distinct architectures in different roles. However, constructing all attention modules in the same network and combining them is time-consuming and inefficient. The neural architecture search (NAS) method has been proposed in \cite{liu2021automatic}, to enhance efficiency and reduce the difficulty in hyperparameter modification. Different attention modules in the same network are searched simultaneously, allowing the network to automatically identify appropriate attention module architectural combinations. 
\subsection{CNN and NLP techniques}
CNN based techniques have been previously used in classifying the PTB-XL dataset. In \cite{e23091121}, the authors have implemented three CNN based techniques on the PTB-XL dataset-(a) based on the convolutional network (b) based on SincNet and (c) based on convolutional network, but with additional entropy-based features, yielding an accuracy of 72\%, 73\% and 76.5\% respectively for classifying five classes. In \cite{s22030904}, the authors have carried out few-shot learning on the PTB-XL dataset achieving the highest accuracy of 80\%. In \cite{SHI2024106253}, the authors have proposed a novel self-supervised learning that fuses generative learning and contrastive learning for denoising and classification of ECG data. The authors reported an accuracy of 78\% on the PTB-XL dataset. In \cite{TAO2024121497}, the authors have proposed a  flexible double-kernel residual block (DKR-block) to extract intra-lead and inter-lead features along with a visualisation method based on gradient-weighted class activation mapping for ECG classification and have achieved an accuracy score of 89.2\%.\\
In \cite{9679174}, the authors have used Beat-aligned Transformer (BaT), where each beat representation is learnt by shifted-window-based Transformer blocks (SW Block), and aggregation blocks are designed to exchange information among the beat representations. This approach on the PTB-XL database has yielded an AUC score of 90\%.

In \cite{song2024bidirectional}, Bidirectional Timely Generative Pre-trained Transformer (BiTimelyGPT), which pre-trains on the time-series PTB-XL dataset by previous token and next token prediction son alternate transformer layers and have achieved an accuracy of 87\% in classifying the ecg signals. 
In \cite{10488393}, the authors have put forward a novel ECG interpretation generation model, where the modifications were introduced in the transformer model to handle 3D input data and generate the spatial information from the same. Even though the study reported an average F1-score of 91\%, the model was computationally expensive and quite challenging to deploy on embedded edge applications. 

\section{Dataset}
\label{dataset}
\begin{figure}
\centering
%\captionsetup{justification=centering,margin=5cm}
\includegraphics[width=0.35\textwidth]{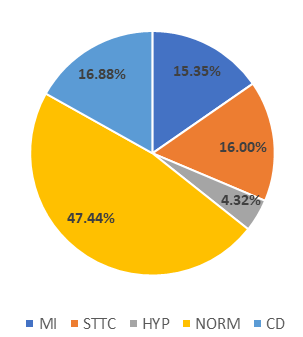}
\caption{Data distribution in ptb-xl dataset}
\label{ptb}
\end{figure}

In this work, publicly available PTB-XL dataset \cite{Wagner2020-xx},\cite{Wagner2022-sy}, derived from Physionet\cite{Goldberger2000-vv}, was used. It is the most comprehensive collection of clinical ECG data. It offers an extensive set of ECG annotations and extra metadata, making it a suitable source for training and assessing machine learning algorithms. The PTB-XL dataset contains 21,837 records of 12-lead 10-sec ECGs from 18,885 individual patients. ECG files are available at 500 Hz and 100 Hz sampling speeds with 16-bit resolution. ECGs with sample rates of 500 Hz were utilised in this study. The database includes 71 different types of cardiac illnesses, divided into five distinct categories: normal ECG (NORM), myocardial infarction (CD), ST/T change (STTC), conduction disruption (MI), and hypertrophy (HYP). Out of 21,837 records, the entries which did not have proper labels assigned to them were removed. In this manner, 17,232 records were utilized in this study, each of which belonged to one of the five classes as shown in Figure \ref{ptb}.

The PTB-XL dataset exhibits a significant imbalance, with almost 48\% of the data belonging to the normal class, while the remaining data is unevenly divided among the four disease classes. Multi-class classification efficacy may be compromised in a number of ways due to data imbalance. Loss function optimization is a component of training a machine learning algorithm where the model incurs a penalty when an incorrect prediction is generated, with each class being penalized equally by the loss function. The majority class exerts a dominant influence on the loss function within an imbalanced dataset, thereby instructing the model to accurately classify the majority class exclusively. This is a critical error in medical AI, where detection of the classes present in minority is of utmost importance. Therefore, to ensure an equitable distribution across all classes, samples are generated for data pertaining to the minority classes.  Synthetic Minority Over-sampling Technique or SMOTE is a powerful technique for addressing data imbalance in multi-class classification problems\cite{10.5555/1622407.1622416}. This technique was employed in this study to ensure a balanced distribution of classes within the data.

\section{Proposed Methodology}
\label{method}
\subsection{Data Preprocessing}
\begin{figure}[htb!]
\centering
%\captionsetup{justification=centering,margin=2cm}
\includegraphics[width=0.5\textwidth]{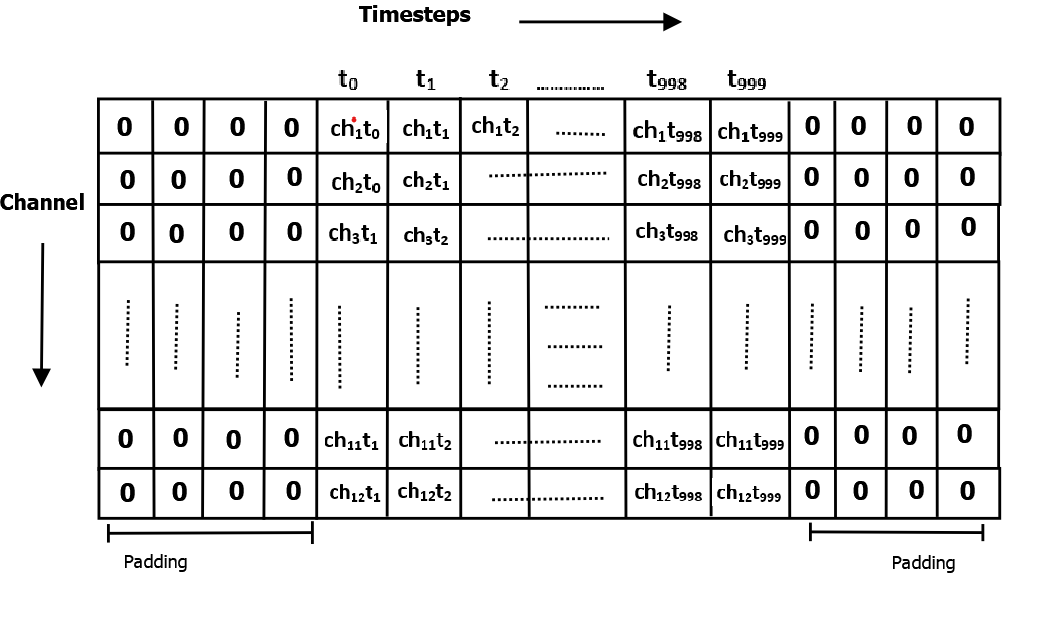}
\caption{Data for a single patient represented as a time series data along with padding}
\label{fig_3}
\end{figure}

Fluctuations in heartbeat are prevalent even among healthy people. As a result, analysing the signal over time reveals anomalies in the heart's activity. The 12-channel ECG data for each patient, was rearranged to effectively represent the temporal relationship of the data across all channels. To prevent loss of data at any given time stamp, zeros were added at the beginning as well as the end of each time series data. The data arrangement along with the padding for a single patient can be visualised in Figure \ref{fig_3}. The figure displays the time series data from each of the twelve channels for a single patient.  

\begin{figure}
\centering
\includegraphics[width=0.5\textwidth]{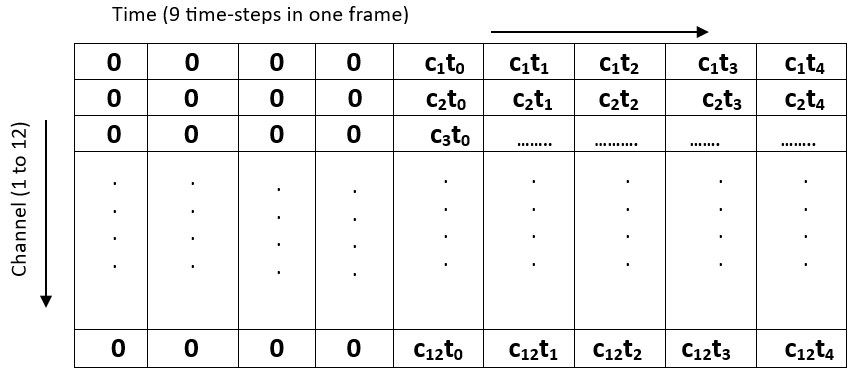} \\
 \vspace{1cm}
\includegraphics[width=0.5\textwidth]{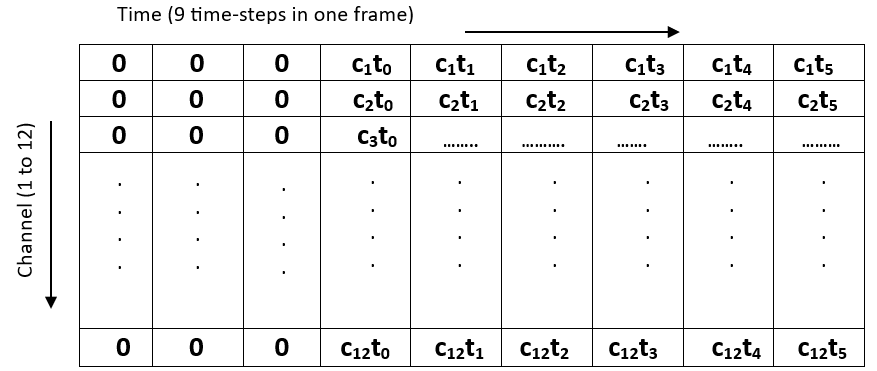}
\caption{First and second set of data for 12 channels of 1 patient after windowing}
\label{fig_4}
\end{figure}

A suitable window of 9 time steps is chosen and the time series data for each patient is split into individual frames with the chosen window size as shown in Figure \ref{fig_4}. The frame generation continues until the window hits the zero-padding, which signifies that the data has ended. Hence, several data frames of 9 time steps are generated for each individual patient.  Figure \ref{fig_4} shows the first and the second frames generated  for a single patient after the windowing operation.

\begin{figure}[htb!]
\centering
%\captionsetup{justification=centering,margin=2cm}
\includegraphics[width=0.5\textwidth]{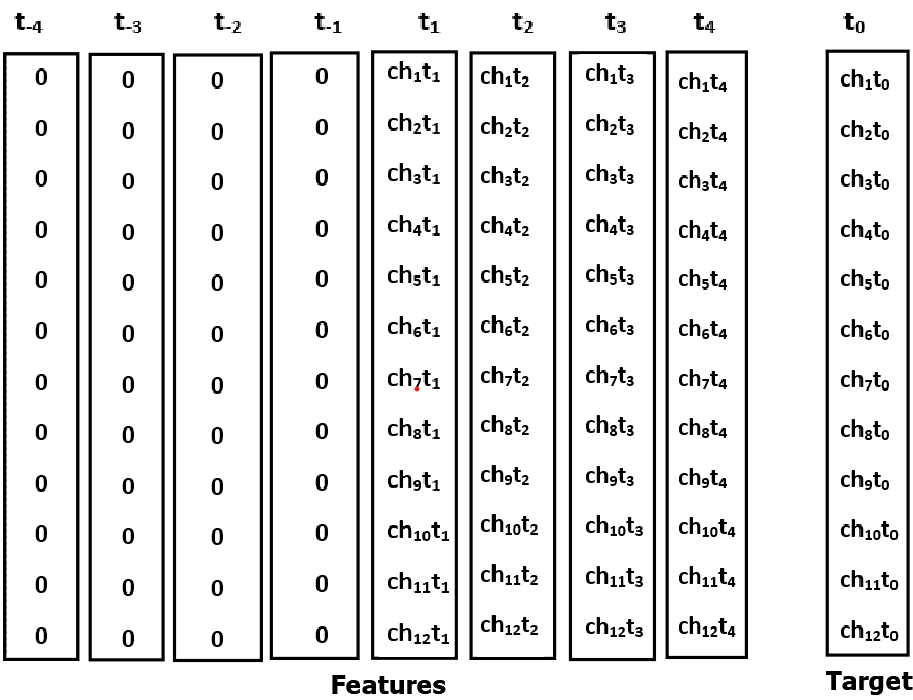} % Just stack two 
\includegraphics[width=0.5\textwidth]{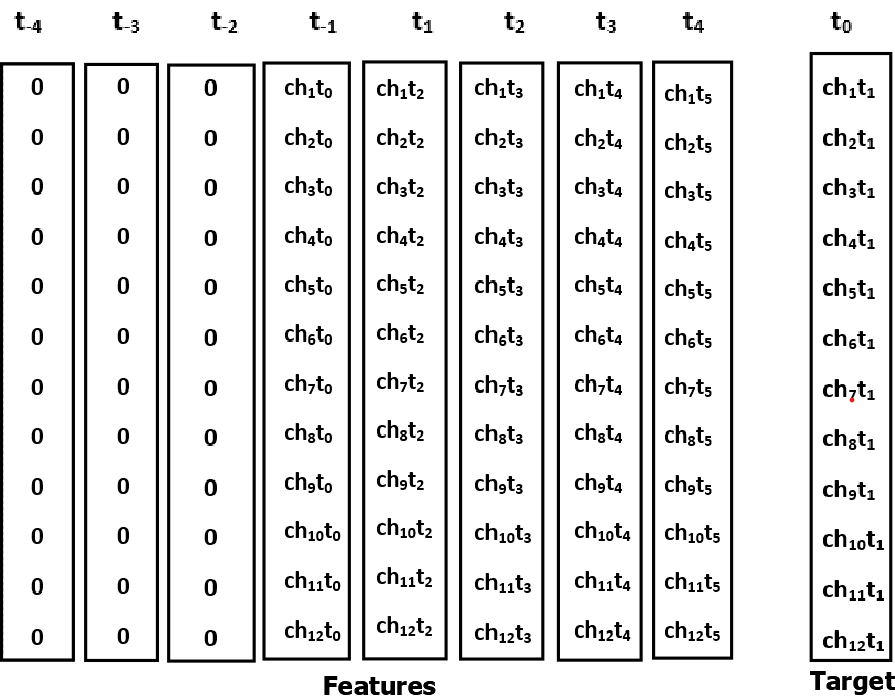}
\caption{Data for a single patient represented as a time series data along with padding}
\label{fig_5}
\end{figure}

Next, the middle column from each frame is split as a separate vector while the remaining four values on the left side as well as the remaining four values on the right side of each frame, are concatenated in the way shown in Figure \ref{fig_5}. Hence, the frames excluding the mid values serves as the feature vector while the vector containing the mid values is the target vector.  Since we have chosen a window of dimension 9 with a padding of 4, the first value of ECG data for each channel is the target vector for the first frame. Subsequently, each time stamp is being assigned as the target vector, while the neighbouring time stamps are being utilised as the feature vector. This feature vector as well as the target vector serves as the training input to the following Temporal Autoencoder Model(TAE) for Data Compression(TAE) network. 

\subsection{Temporal Autoencoder Model(TAE) for Data Compression}
\begin{figure*}[htb!]
%\centering
%\captionsetup{justification=centering,margin=2cm}
\includegraphics[width=7in]{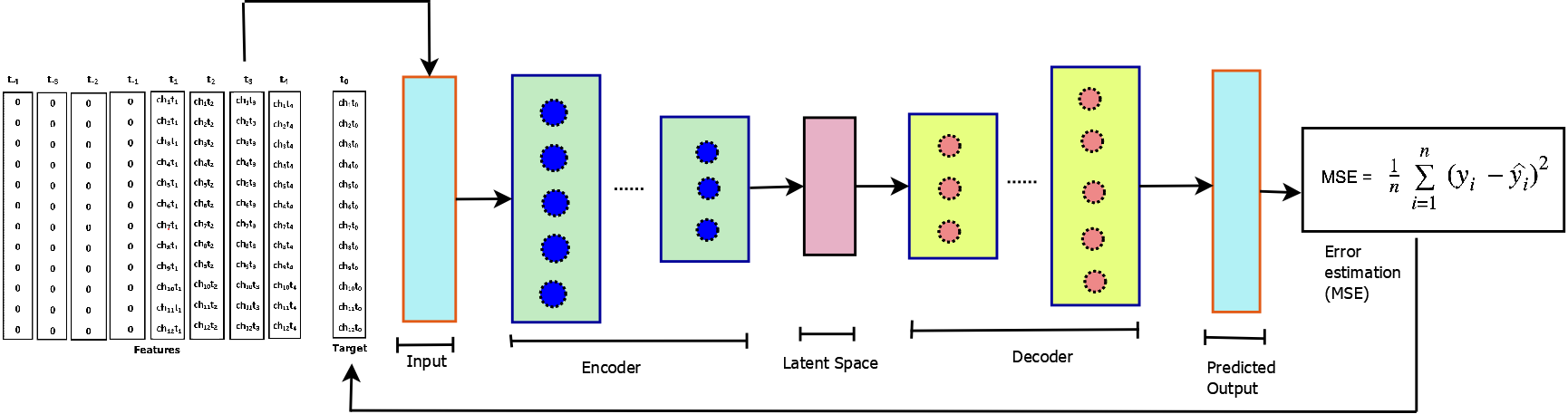}
\caption{Temporal Autoencoder(TAE) Network for Data Compression}
\label{fig_2}
\end{figure*}

\begin{figure*}[!h]
\centering
%\captionsetup{justification=centering,margin=2cm}
\includegraphics[width=7in]{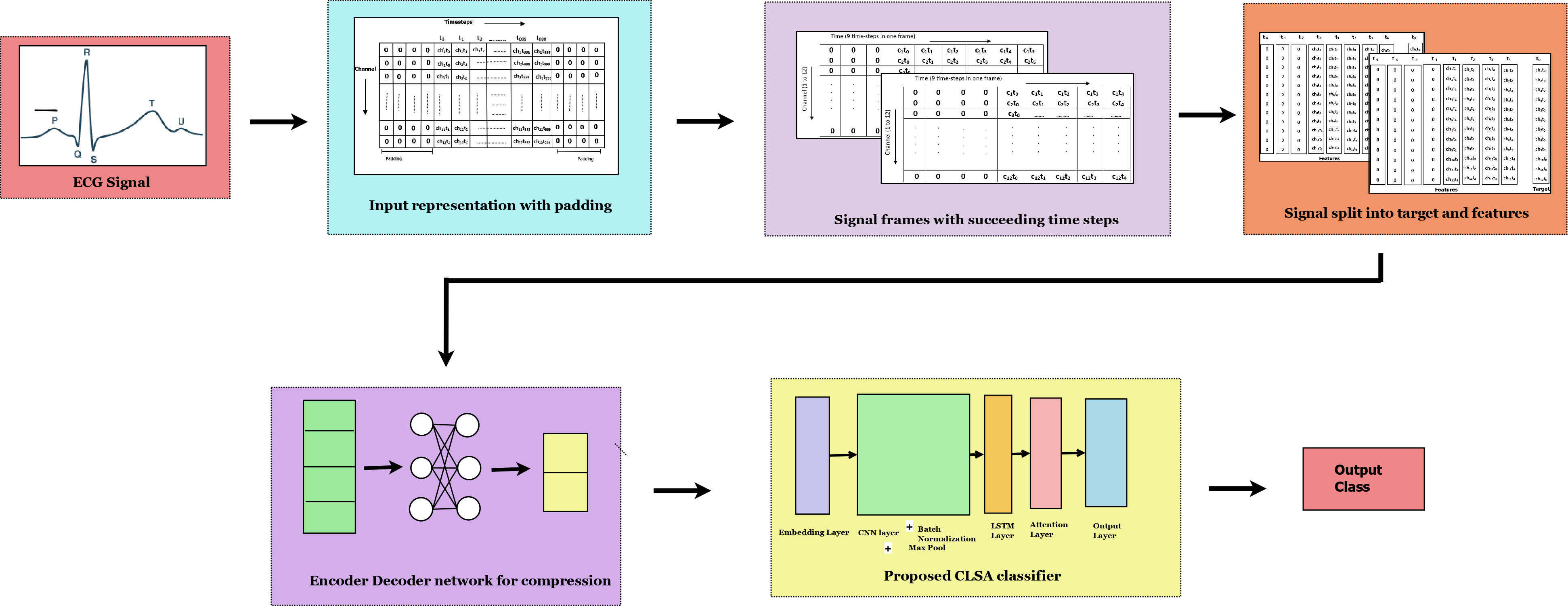}
\caption{Workflow depicting the whole process of data compression and classification}
\label{fig_workf}
\end{figure*}

The architecture of the Temporal Autoencoder Model(TAE) model is elaborated in Figure \ref{fig_2}. The network comprises of  an input layer with 12 nodes, to accommodate the 12 channel input data described in the previous section. Two subsequent dense layers follow, along with a bottleneck/latent layer which serve as the temporal encoder network. The temporal encoder network works on the input data by compressing the same, while trying to capture the most significant features in the data. Two dense layers follow the bottleneck layer which acts as the decoder network. The compressed form of the data, also known as the latent representation is fed to the decoder network which functions as a decompressor, attempting to reconstruct the original data from the compressed form. Finally, we have the output layer from where we obtain the reconstructed output. The details of the TAE network along is shown in Algorithm \ref{algo_1}.  The feature vectors consisting of 12 channels along with the target vectors are fed to the TAE network.  The network is trained by continously comparing the recreated data with the original input data, thereby increasing its capacity to compress and reconstruct the data more efficiently. Over time, it learns to compress the data while maintaining the necessary features for reconstruction. Once the network is trained, the compressed output obtained from the temporal encoder network serves as the input to the embedding layer of the CLSA network described in the next section. Figure \ref{fig_workf} illustrates a comprehensive synopsis of the entire data compression and classification procedure.

\begin{algorithm}
    \caption{Temporal Autoencoder Model(TAE) model}\label{algo_1}
    \begin{algorithmic}
               \REQUIRE TIME SERIES ECG SIGNAL 
               \ENSURE COMPRESSED ECG SIGNAL DATA
               \STATE $1$: Create encoder network with 2 subsequent dense layers with  12 and 8 nodes respectively
               \STATE $2$: Create bottleneck layer with latent dimension of 6
               \STATE $3$: Create decoder network with two subsequent layers with nodes 8 and 12 respectively
               \STATE $4$: Reconstructed ECG signal obtained from output layer with 12 nodes
               \STATE $5$: Compressed ECG representation of dimension 6 obtained from bottleneck layer
                                      
    \end{algorithmic}
\end{algorithm}

% Changes by Wazib STARTS
\subsection{Input Representation}
In this paper, we attempt to represent the ECG signals similar to text representation in Natural Language Processing. For this, each patient's ECG signal is taken as a sequence containing temporal information. Each signal contains 12 channels comprising 1000 temporal instances. Analogous to input representation in NLP, each patient's sample is treated as a sequence. The 1000 temporal instances are treated as tokens in the sequence having 12 embedding components corresponding to each token. This is similar to the input fed into a conventional NLP model with each word in the $i^{th}$ sequence $Seq_{i}$ converted to its equivalent embedded representation $e_j$ having $D$ components.

\begin{equation}Seq={e_j^d \textit{, } \forall d \in D \textit{ and } \forall j \in |V|}  \label{sequence}\end{equation}

where, the total number of unique embeddings equals $|V|$, i.e. the size of the vocabulary $V$. Here, each unique word $w_{j}$ in $V$ is denoted by its corresponding embedding $e_j$. For this, a lookup table is constructed containing the mapping between the words and their embeddings.

For a particular patient $i$, their ECG sequence representation $s_{i}$ is as follows: 

\begin{equation}s_{i}=[[t^{0}_{0}, t^{0}_{1}, ... , t^{0}_{12}], [t^{1}_{0}, t^{1}_{1}, ... , t^{1}_{12}], ... , [t^{999}_{0}, t^{999}_{1}, ... , t^{999}_{12}]] \label{seq_1}\end{equation}

\par for $\textit{, } s_i \in S$ \\

where, $t^{k}_{j}$ denotes the $j^{th}$ temporal instance value belonging to $k^{th}$ channel, and $S$ is the entire corpus containing $n$ sequences equal to the total number of patient records. To represent this sequence in terms of the representation followed in NLP, we identify each unique temporal instance $t_j$ in the total corpus and prepare a vocabulary $V'$ out of them along with a lookup table where $t_j$ is mapped as follows:
\newline

\begin{equation}t_j= et_j^c \textit{, } \forall c \in C\label{lookup}\end{equation}
\newline

where, $C$ denotes the number of ECG channels. 

\subsection{Embedding Approach}

An embedding layer can be conceptualized as a transformation that translates words into their corresponding word embeddings. An embedding layer receives a sequence of numbers, imitating a sequence of words in a sentence, as input. Within the embedding layer, is the embedding matrix. It is a two-dimensional array in which each row corresponds to a distinct categorical input and each column contains a vector of continuous values. These vectors containing continuous values are sometimes known as "embeddings."
Let the sequence of discrete categorical values which serves as the inputs to the embedding layer be denoted by \textbf{$x$}. Let \textbf{$V$} be the size of the vocabulary, containing \textbf{$V$} total unique embeddings. Each input should be an embedding vector of dimension \textbf{$D$}. Hence, \textbf{$x$} can take values ranging from 0 to \textbf{$D-1$}. Hence, the dimension of the embedding matrix is \textbf{($V$, $D$)}. The output of the embedding layer is denoted by \textbf{$e$($x$)}.

\subsection{Convolutional Neural Networks(CNN)}
CNNs are inherently designed for grid-like data, such as images, but have transformed into a rapid and effective solution for various natural language processing tasks. A CNN carries out a sequence of convolution and pooling operations, subsequently transitioning to a fully connected neural network. These models are adept at identifying patterns within the input, regardless of their spatial location. As these layers are iterated, these patterns can be recognized at more abstract levels, ultimately facilitating the identification of significant data features.
\par In this work, 1D-CNN has been used on the output obtained from the preceding embedding layer. This layer extracts local features from the signals which include spikes, waves, and other patterns which might be indicative of a worsening heart condition.

\subsection{Long Short-Term Memory(LSTM)}
 Long Short-Term Memory(LSTM) is a type of recurrent neural network (RNN) architecture that is used particularly for tasks that involve sequential data. They are capable of capturing long-range dependencies in sequences, making them well-suited for tasks where understanding context over extended periods is important. LSTMs are effective for processing sequential data, such as time series data as they can capture the temporal dependencies and can identify unusual patterns or behaviors in a sequence of events. 

\subsection {Self Attention}
Self-attention has demonstrated its effectiveness in capturing long-range dependencies. In self-attention, the goal is to compute a weighted sum of input values for each element in a sequence, where the weights are determined based on the similarity or relevance of elements in the sequence to a given element. This mechanism is particularly useful for capturing relationships and dependencies between elements in a sequence. If a sequence of input vectors is denoted as X, where X = [$x_1$, $x_2$, ..., $x_n$], and each $x_i$ represents an element in the sequence. These input vectors can be thought of as the raw features or representations of the sequence.
\par To compute key (K), query (Q), and value (V) vectors, linear transformations are applied to the input vectors:\\

\begin{equation}Key (K): K = XW\_K \label{key}\end{equation}
\begin{equation}Query (Q): Q = XW\_Q \label{key}\end{equation}
\begin{equation}Value (V): V = XW\_V  \label{key}\end{equation}

\par Here, W\_K, W\_Q, and W\_V are learnable weight matrices. These transformations allow the network to learn which parts of the input are important as keys, which parts are relevant as queries, and which parts contain information to be retrieved as values. 
\par In self-attention, each element in the sequence serves as both a query and a key. This means that the queries and keys are identical, allowing each element to assess its own relationship with all other elements in the sequence. For self-attention, the similarity scores are calculated for each pair of elements in the sequence. Since query and key are the same for each element, this can be represented as a dot product between the input vectors:\\

\begin{equation}Similarity scores:S=X.X^T\label{sim-score}\end{equation}

Here, S is a matrix where each entry S(i, j) represents the similarity between the i-th element and the j-th element. This means that each element assesses its own similarity to all other elements in the sequence. The similarity scores are scaled and the softmax function is applied to obtain the attention weights.

\begin{equation} Scaled Similarity scores: S'= S / sqrt(d)\label{sim-score_2}\end{equation}, 
\par where "d" is the dimension of the input vectors\\

\begin{equation} Attention Weights: A = softmax(S')\label{att}\end{equation}

Finally, the attention weights are used to compute the weighted sum of the input vectors:\\

\begin{equation} Weighted Sum: Z = AX\label{att}\end{equation}

\par Z is the output of the self-attention mechanism, representing the attended features. Each element in Z corresponds to a weighted sum of all input vectors, where the weights are determined by the attention distribution for the corresponding element.

\subsection{ CLSA(CNN-LSTM-Self Attention) model}

\begin{figure*}[htb!]
\centering
%\captionsetup{justification=centering}
\includegraphics[width=6in]{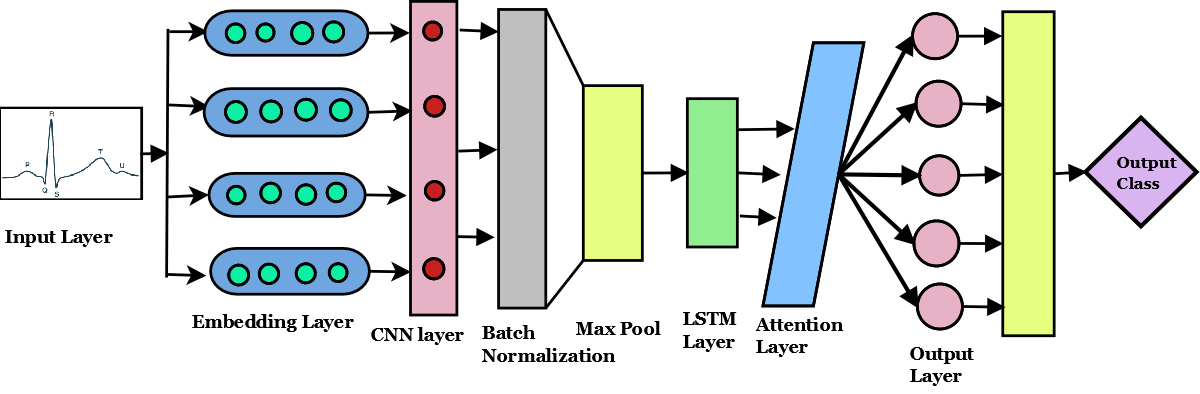}
\caption{CLSA Network Architecture}
\label{fig_1}
\end{figure*}

One may think of the input layer as 1000 tokens, with 6 embedding components per token, corresponding to the 6-dimensional output obtained from the temporal encoder network, for each patient in the dataset.  This layer serves as the embedding layer's input, while the layer's output is an embedding matrix with dimensions (V,D). The compressed output obtained from the temporal encoder network is passed on to the embedding layer as weights. The 1-D CNN layer, takes this embedding matrix as the input. The convolution operation involves computing the dot product between a filter and a localized section of the input embeddings. This process is executed throughout the entire input sequence, resulting in the creation of a set of feature maps. CNN is followed by a Batch-normalization which normalizes the input by subtracting the mean and dividing by the standard deviation of the mini-batch. Batch normalisation stabilizes the learning process which improves the performance of neural networks by minimising internal covariate shift and enhancing gradient flow. Next,  a Pooling layer was incorporated into the network to downsample the spatial dimension of the data. This reduces computational complexity along with minimizing the risk of overfitting in the model. 
\\
Subsequently, an LSTM layer follows, which analyzes a sequence of data by iterating through time steps and learns the long-term dependencies between them  by producing hidden states. The hidden states from the LSTM are fed into the following self-attention layer. The self-attention mechanism treats each hidden state as a query, key, and value. Attention scores, which show how important each component is in the sequence in relation to the others, are calculated by the self-attention layer. This weighted total, also known as the context vector, gathers pertinent data from various segments of the sequence. A couple of fully connected layers are incorporated into the model as the penultimate layers. Finally, we have the output layer consisting of five output nodes corresponding to the five classes in the dataset. Softmax is used as the activation function which indicates the probability of the sample belonging to any particular class. Figure \ref{fig_1} gives a pictorial representation of the proposed novel classification model described in this section.

The hyperparameters used in this network are elaborated in detail in section \ref{hyper}, while the results obtained using the proposed classifier have been demonstrated in section \ref{results}

\begin{algorithm}
    \caption{CNN-LSTM-Self Attention model}\label{algo}
    \begin{algorithmic}
               \REQUIRE TIME SERIES ECG SIGNAL 
               \ENSURE OUTPUT CLASS INDICATING NORMAL/DISEASE
               \STATE $1$: Create embedding layer with the weights obtained by the autoencoder model
               \STATE $2$: Perform 1D Convolution with 3x3 kernel
               \STATE $3$: Perform 20\% Dropout 
               \STATE $4$: Hidden representation of CNN passed on to LSTM with 20\% Kernel regularizer
               \STATE $5$: Compute self attention on the hidden representation from LSTM
               \STATE $6$: 1st Dense layer with ReLU activation
               \STATE $7$: 2nd Dense layer with ReLU
               \STATE $8$: Final dense layer with output nodes and sigmoid activation function                                            
    \end{algorithmic}
\end{algorithm}

\begin{comment}
    \subsection{Citations to the Bibliography}
The coding for the citations is made with the \LaTeX\ $\backslash${\tt{cite}} command. 
This will display as: see \cite{ref1}.

For multiple citations code as follows: {\tt{$\backslash$cite\{ref1,ref2,ref3\}}}
 which will produce \cite{ref1,ref2,ref3}. For reference ranges that are not consecutive code as {\tt{$\backslash$cite\{ref1,ref2,ref3,ref9\}}} which will produce  \cite{ref1,ref2,ref3,ref9}

\end{comment}

\section{Experimental Setup and Hyperparameter details }
\label{hyper}

 \par In this study, the normal class, which represents the majority, consists of 9083 samples, whereas the remaining classes collectively accounted for 8149 samples. To address the significant disparity in data, the SMOTE oversampling technique was employed, resulting in all data samples being adjusted to a value of 9083. Consequently, the total number of data samples amounted to 45415. 

Additionally, to compress the data, feature vectors were created for each of the 1000 time stamps in the dataset. This resulted in the total dataset of 45415 samples being multiplied by 1000. The resultant dataset exceeded the capacity of the TAE network to be trained in a single batch. Hence, the dataset was divided into two equal batches. The dataset for each batch was partitioned into three subsets: training, validation, and testing, with a distribution ratio of 80-10-10, respectively. The model was trained using the first segment of the data. The trained model, along with its learned parameters and structure, was stored. This model was revoked when training with the second batch of the data. This two-step training had the added advantage of optimizing the model parameters by iterating the model twice. Table \ref{tab:Enc-dec} lists the dimensions and activation functions employed in the TAE network.
 
\begin{table}[htb]
\centering
\caption{Hyperparameters used in Encoder-decoder network}
\vspace{5pt}

\begin{tabular}{|c | c | c | c |} 
 \hline
\textbf{Layer} & \textbf{Type} & \textbf{No.of Units} & \textbf{Activation}\\  
 \hline
 \hline
Input  & & 12 & -\\
 \hline
Encoder$_1$ & Dense & 12  & ReLU\\
 \hline
 Encoder$_2$ & Dense & 8  & ReLU\\
 \hline
Latent input & Dense & 6 & Linear\\
 \hline
 Decoder$_1$ & Dense & 8 & ReLU  \\
\hline 
Decoder$_2$ & Dense & 12 & ReLU  \\
\hline
Output & & 12 & ReLU\\
\hline
\end{tabular}
\label{tab:Enc-dec}
\end{table}

The compressed inputs obtained from the hidden layer is fed to the proposed CLSA classifier. The dataset was partitioned into three subsets: training, validation, and testing, with a distribution ratio of 80-10-10, respectively. The hyperparameters used in the CLSA network have been elaborated in Table \ref{tab:hyp_clsa}.

\begin{table}[htb]
\centering
\caption{Hyperparameters used in CLSA network}
\vspace{5pt}

\begin{tabular}{|c | c | c | c |} 
 \hline
\textbf{Layer} & \textbf{No.of Units} & \textbf{Kernel Size} & \textbf{Activation}\\  
 \hline
 \hline
 CNN & 512 & 3  & ReLU\\
 \hline
 LSTM & 256 &  -- & Tanh\\
 \hline
 L2 Regularization & 20\% & -- &\\
 \hline
 Max. Pool & -- &2 & --  \\
\hline 
Dense & 64 & -- & ReLU \\
\hline
Dense & 32 & -- & ReLU\\
\hline
Dense(Output) & 5 & -- & Softmax\\
\hline
\end{tabular}
\label{tab:hyp_clsa}
\end{table}

\section{Results}
\label{results}
 In this section, the ECG classification results obtained using the proposed methodology have been presented. Firstly, the outcomes of the signal compression technique using the proposed TAE network have been discussed. Secondly, the effectiveness of using the self-attention mechanism, to classify the ECG signals, has been presented.

 \subsection{Results from TAE network}
\begin{figure}
\centering
%\captionsetup{justification=centering,margin=2cm}
\includegraphics[width=3in]{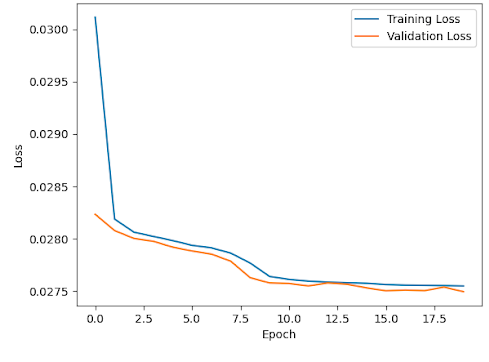}
\caption{Training vs Validation loss plot for the first batch of encoder-decoder}
\label{auto_enc_1}
\end{figure}

\begin{figure}
\centering
%\captionsetup{justification=centering,margin=2cm}
\includegraphics[width=3in]{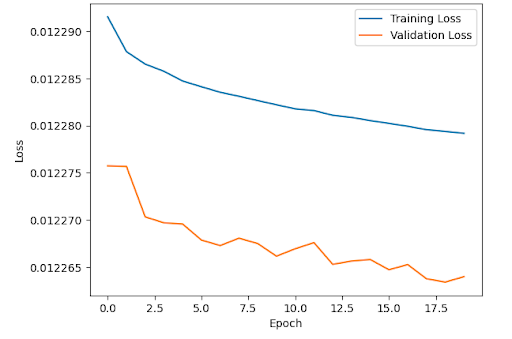}
\caption{Training vs Validation loss plot for the second batch
of encoder-decoder}
\label{auto_enc_2}
\end{figure}

 The loss function selected for assessing the performance of the model is the "Mean Square Error"(MSE). The MSE algorithm calculates the squared difference between each element in the predicted output vector and the corresponding element in the ground truth. It then computes the average of these squared differences across all elements. This loss is commonly utilized with continuous data and is recognized for its susceptibility to outliers. A low value of MSE indicates a better fit of the model to the data, i.e., the closer the predicted values are to the actual values.
Figure \ref{auto_enc_1} and Figure \ref{auto_enc_2} illustrate the training vs. validation loss for the 1st and the 2nd batch of the encoder-decoder model respectively.
The mean square errors for the 1st and the 2nd batches of the models were calculated and found to be 0.023 and 0.0124 respectively. Hence, it is demonstrated that the model gets optimized for improved efficiency and accurate data replication by getting trained twice. The bottleneck layer of the encoder model yielded a latent representation of the data consisting of 6 channels. This data is used as the input for our proposed CLSA network. 

\begin{table}[htb!]
\centering
\caption{Performance metrics of Autoencoder network}
\vspace{5pt}

\begin{tabular}{|c | c |} 
 \hline
 \hline
Autoencoder Batch & Mean Square Error Loss\\  
 \hline
 \hline
1 & 0.023\\
\hline
2 & 0.0124\\
 \hline
 \end{tabular}
\label{tab:multiclass_score}
\end{table}

\subsection{Results from CLSA network}

Accuracy is a metric used to evaluate the performance of a classification model. It is defined as the ratio of correctly predicted instances to the total number of instances. The formula for accuracy is:\\

$Accuracy = \frac{TP+TN}{TP+TN+FP+FN}$\\

Precision measures the accuracy of the positive predictions made by the model. Precision is defined as the ratio of true positive predictions to the total number of positive predictions (both true positives and false positives). High precision indicates that the model has a low false positive rate, meaning it is good at predicting the positive class accurately. The formula for Precision is: \\

$Precision = \frac{TP}{TP+FP}$\\

Recall measures the proportion of actual positives that are correctly identified by the model. Recall is defined as the ratio of true positive predictions to the total number of actual positives (the sum of true positives and false negatives). High recall indicates that the model is effective at identifying the positive class, with fewer positive instances being missed. The formula for recall is:\\

$Recall = \frac{TP}{TP+FN}$\\

F1-score is the harmonic mean of precision and recall, providing a single measure that balances both concerns. A good F1-score reflects the model's robustness in making accurate predictions across all classes.The F1 score is defined as follows:\\\\
$F1 = \frac{2*Precision*Recall}{Precision+Recall} = \frac{2*TP}{2*TP+FP+FN}$\\

In the above equations, 
\begin{itemize}
    \item TP is the number of true positives
    \item TN is the number of true negatives
    \item FP is the number of false positives
    \item FN is the number of false negatives
\end{itemize}

\begin{figure}
\centering
%\captionsetup{justification=centering,margin=2cm}
\includegraphics[width=3.5in, height=2in]{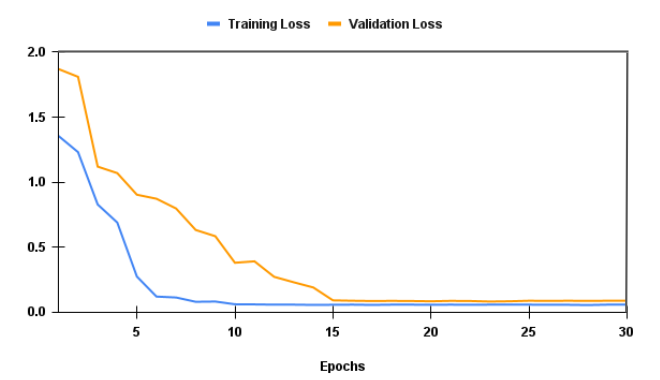}
\caption{Training vs Validation loss plot across 30 epochs for training the CLSA network}
\label{clsa_1}
\end{figure}

\begin{figure}
\centering
%\captionsetup{justification=centering,margin=2cm}
\includegraphics[width=3.5in, height=2in]{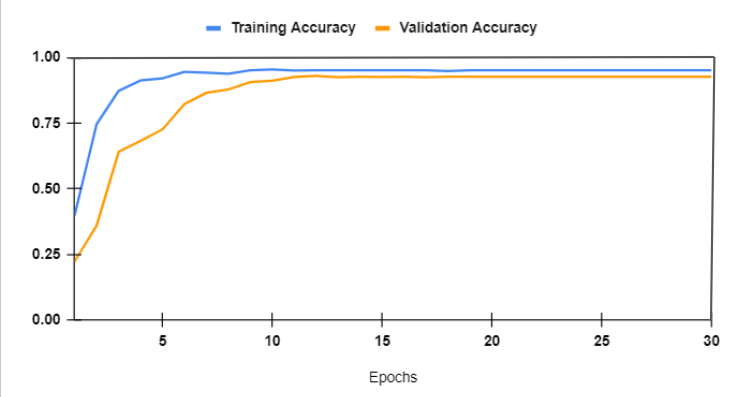}
\caption{Training vs Validation accuracy across 30 epochs for training the CLSA network}
\label{clsa_2}
\end{figure}

A classification accuracy of 91\% is achieved with the proposed TAE-CLSA model, which indicates that the model is effective in identifying the correct classes during classification of ECG signals. Figure\ref{clsa_1} shows the variation of the training loss vs validation loss while training the CLSA network across 30 epochs while Figure\ref{clsa_2} shows the variation of the training vs validation accuracy across the 30 epochs.
 An F1-score of more than 85\%  is obtained for all the classes as has been elaborated in Table \ref{tab:multiclass_score}. An F1 score of 85\% and above indicates that the model is effective in identifying the True Positives while keeping the keeping the False Positives and False Negatives low. This is particularly important in cases which deals with medical data where an incorrect prediction of False Positive or False Negative may lead to a fatal result. 

\par A comparison of our work with existing works has been presented in Table \ref{table:comparison}. It can be seen that the F1 scores achieved using our proposed method have surpassed the results of the existing works in most of the cases, to the best of our knowledge. 

\begin{table}[htb!]
\centering
\caption{Performance metrics of CLSA classifier}
\vspace{5pt}

\begin{tabular}{|c c c c|} 
 \hline
 \hline
Class & Sensitivity & Precision & F1-score \\  
 \hline
 \hline
CD & 0.88 & 0.93 & 0.90  \\ 
 \hline
 HYP & 0.94 & 0.95 & 0.95 \\
 \hline
 MI & 0.84 & 0.86 & 0.85 \\
 \hline
 NORM & 0.95 & 0.87 & 0.88 \\
 \hline
 STTC & 0.86 & 0.87 & 0.86 \\
 \hline
 & Accuracy=91\% &&\\
 \hline
 \end{tabular}
\label{tab:multiclass_score}
\end{table}

\begin{table*}
%\begin{minipage}{\textwidth}  
\begin{center}
%\vspace{-6pt}
\caption{Comparison of our proposed method with the existing works}
\scalebox{0.8}{
\begin{tabular}{|c|c|c|c|c|c|c|c|c|c|c|c|c|c|c|}
\hline 
%\cline{2-7}
%\hline
& & \multicolumn{2}{|c|}{CD}&\multicolumn{2}{|c|}{HYP}&\multicolumn{2}{|c|}{MI}&\multicolumn{2}{|c|}{NORM}&\multicolumn{2}{|c|}{STTC} & \\
\cline{3-12}
Classifier & Algorithm & Se & P$_{+}$ & Se &P$_{+}$ & Se & P$_{+}$ & Se &P$_{+}$ &  Se & P$_{+}$ & {Accuracy}\\
\hline
Palczynski et.al.\cite{palczynski2022study} & CNN+few shot learning & 72.2 & 65.2 & 55.9 & 47.27 & 72.8 & 74.01 & 87.9 & 89.7 & 63.8 & 69.2 & 77.18 \\
\hline
Smigiel et.al\cite{e23091121} & CNN+entropy features & 79.78 & 59.56 & 60.56 & 39.45 & 69.85 & 76.69 & 83.35 & 93.10 & 63.7 &  62.6 & 76.52\\
\hline
Virgeniya et al.\cite{S2021102779}& GRU-ELM &  97.4 & 89 & 99.9 & 95.5 & 98.5 & 88.7 & 71.5 & 77.4 & 96.7 & 87.1 & 87.3\\
\hline
Jyotishi et.al\cite{10080902}& Attentive spatio-temporal
learning-based neural network & 76.6 & 76.46 &  54.19 & 52.6 & 73.62 & 75.5 & 80.5 & 92.96 & 74.53  & 80.5 & 85.5\\
\hline
Geng et.al\cite{healthcare11071000} & SE-ResNet + Contextual transformer(CoT) & 85.9 & 86 & 85 & 78.9 & 76.6 & 74.9 & 85.5 & 83.2 & 83.5 & 81 &87.9 \\
 \hline
Proposed method & Conv+lstm+attention &  88 & \textbf{93} & 94 & \textbf{95} & 84 & 86 & \textbf{95} & 87 & 86 & \textbf{87} & \textbf{91} \\
\hline
\multicolumn{4}{l}{\small *Se indicates sensitivity or recall  and P$_{+}$ indicates precision.} \\
\end{tabular}}
\label{table:comparison}
%\vspace{-18pt}
\end{center}
\end{table*}

\section{Ablation Study}
\label{ablation}
In this study we present the effect of variation in the hyperparameters as well as several other parameters of the network which well justifies the choice of the hyperparameters used in the proposed network. The values highlighted in bold indicate the values used in this study.

\subsection{Effect of using TAE model}
In the first step, the performance of the model devoid of the autoencoder network was studied. No data compression was performed and hence, a 12 channel data was fit to the network, the results of which has been illustrated in Table \ref{tab:abl_TAE}.

\begin{table}[h]
\centering
\caption{Effect of using TAE network with CLSA classifier}
\vspace{5pt}

\begin{tabular}{|c | c | c |} 
 \hline
\textbf{Model used} & \textbf{Learning Rate} & \textbf{Inference Accuracy} \\
\hline
\hline
Only CLSA & 0.1 & 85.1\%\\
\hline
Only CLSA & 0.01 & 85.5\%\\
\hline
Only CLSA & 0.001 & 87\%\\
\hline
CLSA +TAE & 0.1 & 88\%\\
\hline
CLSA + TAE & 0.01 & 90.2\&\\
\hline
\textbf{CLSA + TAE} & \textbf{0.001} & \textbf{91}\%\\ 
\hline
\end{tabular}
\label{tab:abl_TAE}
\end{table}

\subsection{Effect of variation of CNN and LSTM units on accuracy}
The number of filters in the CNN classifier as well as the LSTM layer was varied to find the optimal setting which would give a decent result. On increasing the number of filters above 512, a marginal improvement in inference accuracy was obtained while the model size as well as complexity was on the rise. Hence, an optimum setting of 512 units for CNN and 256 units for LSTM was chosen. The effect of the number of units convolution units, lstm units and the kernel size is illustrated in Table \ref{tab:abl_CNN}.

\begin{table}[htb]
\centering
\caption{Effect of Model variation on Inference accuracy}
 \setlength{\tabcolsep}{0.7\tabcolsep}% Shrink \tabcolsep by 30%
  \centering
\vspace{5pt}
\begin{tabular}{|c | c | c | c |} 
 \hline
\textbf{No. of Conv. units} & \textbf{Kernel Dim.} & \textbf{No. of LSTM units} & \textbf{Accuracy} \\
\hline
\hline
128 & 3x3 & 128 & 78\%\\
\hline
256 & 3x3 & 128 & 78.6\%\\
\hline
256 & 3x3 & 256 & 85\%\\
\hline
256 & 5x5 & 256 & 85\%\\
\hline
256 & 3x3 & 128 & 85\%\\
\hline
512 & 3x3 & 128 & 87\%\\
\hline
512 & 5x5 & 128 & 88\%\\
\hline
\textbf{512} & \textbf{3x3} & \textbf{256} & \textbf{91}\%\\
\hline
500 & 3x3 & 512 & 91.5\%\\
\hline
1024 & 3x3 & 512 & 91.6\%\\
\hline
1024 & 3x3 & 1024 & 91.2\%\\
\hline
\end{tabular}
\label{tab:abl_CNN}
\end{table}

\subsection{Effect of window size on data compression} 
Window size was another factor which was considered while tuning the model as shown in Table \ref{tab:abl_TAE_win} . While a window size of 9 gave us an accuracy of 91\%, a window size of 11 produced an accuracy of 91.3\%, however the model took almost 1.5 times for inferencing along with a rise in the model size. In case of a smaller window size of 5, the accuracy dropped significantly to 86\%. Hence, a tradeoff was done to set the window size to 9.

\begin{table}[htb!]
\centering
\caption{Effect of window size on inference accuracy}
\vspace{5pt}

\begin{tabular}{|c | c | c |} 
 \hline
\textbf{Window size} & \textbf{Inference Accuracy} \\
\hline
\hline
\hline
11 & 91.8\%\\
\hline
7 & 91.3\%\\
\hline
\textbf{9} & \textbf{91}\%\\
\hline
5 & 86\%\\
\hline
\end{tabular}
\label{tab:abl_TAE_win}
\end{table}

\subsection{Effect of latent dimension for data compression}
Various latent dimensions were taken into account while tuning to model as shown in Table \ref{tab:abl_TAE}. In the present setting a latent dimension of 6 was chosen to obtain an accuracy of 91\%. When the latent dimension was increased to 8 and then 10, there was no significant increase in the training as well as the inference accuracy, with the model achieving 91.7\% with latent dimension 10. However, the model size increased by almost 40\% when the latent dimension was increased to 10. It was not justified to increase the model size for such a small improvement in accuracy. Hence, latent dimension was chosen at 6.

\begin{table}[htb!]
\centering
\caption{Effect of latent dimension on inference accuracy}
\vspace{5pt}

\begin{tabular}{|c | c | c |} 
 \hline
\textbf{Latent Dimension} & \textbf{Inference Accuracy} \\
\hline
\hline
\hline
12 & 91.7\%\\
\hline
10 & 89.6\%\\
\hline
8 & 91.3\%\\
\hline
\textbf{6} & \textbf{91}\%\\
\hline
4 & 87\%\\
\hline
\end{tabular}
\label{tab:abl_TAE_CLSA}
\end{table}

\section{Conclusion}
\label{conclusion}
In this work, a compressed network for ECG classification based on NLP technique was presented. A new technique for signal embedding was proposed where the vocabulary of the signal is not limited beforehand, keeping provisions for inclusion of unseen signals. Self attention was incorporated in the classification model to focus on the important features in the temporal sequence of signals. An accuracy of 91\% was achieved with a good percentage of detection for minority classes too. Additionally, a significant reduction of 31\% model size was achieved due to compression. Hence, we can conclude that the proposed TAE-CLSA technique is reliable for detection of cardiovascular abnormalities and being lightweight can also be deployed on edge devices for portability. Future scope includes extending the model on several other datasets and deploying the model on edge devices to explore the possibility of implementing the algorithm in real time.

\bibliographystyle{unsrt}  
%\bibliography{references}  

\end{document}